\documentclass[dvips]{acta}
\usepackage{supertabular,lscape,epsfig}
\usepackage{amssymb}
\usepackage{amsmath}
\usepackage{placeins}
\DeclareSymbolFont{ppa}{OT1}{ppl}{m}{it}
\DeclareMathSymbol{\vv}{\mathalpha}{ppa}{'166}

\SetPages{1}{1}

\SetVol{68}{2018}

\begin{document}

\newcommand{\TabApp}[2]{\begin{center}\parbox[t]{#1}{\centerline{
  {\bf Appendix}}
  \vskip2mm
  \centerline{\small {\spaceskip 2pt plus 1pt minus 1pt T a b l e}
  \refstepcounter{table}\thetable}
  \vskip2mm
  \centerline{\footnotesize #2}}
  \vskip3mm
\end{center}}

\newcommand{\TabCapp}[2]{\begin{center}\parbox[t]{#1}{\centerline{
  \small {\spaceskip 2pt plus 1pt minus 1pt T a b l e}
  \refstepcounter{table}\thetable}
  \vskip2mm
  \centerline{\footnotesize #2}}
  \vskip3mm
\end{center}}

\newcommand{\TTabCap}[3]{\begin{center}\parbox[t]{#1}{\centerline{
  \small {\spaceskip 2pt plus 1pt minus 1pt T a b l e}
  \refstepcounter{table}\thetable}
  \vskip2mm
  \centerline{\footnotesize #2}
  \centerline{\footnotesize #3}}
  \vskip1mm
\end{center}}

\newcommand{\MakeTableApp}[4]{\begin{table}[p]\TabApp{#2}{#3}
  \begin{center} \TableFont \begin{tabular}{#1} #4 
  \end{tabular}\end{center}\end{table}}

\newcommand{\MakeTableSepp}[4]{\begin{table}[p]\TabCapp{#2}{#3}
  \begin{center} \TableFont \begin{tabular}{#1} #4 
  \end{tabular}\end{center}\end{table}}

\newcommand{\MakeTableee}[4]{\begin{table}[htb]\TabCapp{#2}{#3}
  \begin{center} \TableFont \begin{tabular}{#1} #4
  \end{tabular}\end{center}\end{table}}

\newcommand{\MakeTablee}[5]{\begin{table}[htb]\TTabCap{#2}{#3}{#4}
  \begin{center} \TableFont \begin{tabular}{#1} #5 
  \end{tabular}\end{center}\end{table}}

\newfont{\bb}{ptmbi8t at 12pt}
\newfont{\bbb}{cmbxti10}
\newfont{\bbbb}{cmbxti10 at 9pt}
\newcommand{\uprule}{\rule{0pt}{2.5ex}}
\newcommand{\douprule}{\rule[-2ex]{0pt}{4.5ex}}
\newcommand{\dorule}{\rule[-2ex]{0pt}{2ex}}
\def\thefootnote{\fnsymbol{footnote}}
\begin{Titlepage}

\Title{The OGLE Collection of Variable Stars.\\
Type II Cepheids in the Magellanic System\footnote{Based on observations
obtained with the 1.3-m Warsaw telescope at the Las Campanas Observatory of
the Carnegie Institution for Science.}}
\Author{I.~~S~o~s~z~y~ñ~s~k~i$^1$,~~
A.~~U~d~a~l~s~k~i$^1$,~~
M.\,K.~~S~z~y~m~a~ñ~s~k~i$^1$,~~
\L.~~W~y~r~z~y~k~o~w~s~k~i$^1$,\\
K.~~U~l~a~c~z~y~k$^2$,~~
R.~~P~o~l~e~s~k~i$^3$,~~
P.~~P~i~e~t~r~u~k~o~w~i~c~z$^1$,~~
S.~~K~o~z~³~o~w~s~k~i$^1$,\\
D.~~S~k~o~w~r~o~n$^1$,~~
J.~~S~k~o~w~r~o~n$^1$,~~
P.~~M~r~ó~z$^1$,~~
K.~~R~y~b~i~c~k~i$^1$,
and~~P.~~I~w~a~n~e~k$^1$
}
{$^1$Warsaw University Observatory, Al.~Ujazdowskie~4, 00-478~Warszawa, Poland\\
e-mail: soszynsk@astrouw.edu.pl\\
$^2$Department of Physics, University of Warwick, Gibbet Hill Road, Coventry, CV4~7AL,~UK\\
$^3$Department of Astronomy, Ohio State University, 140 W. 18th Ave., Columbus, OH~43210, USA}
\Received{~}
\end{Titlepage}


\Abstract{We present a nearly complete collection of type~II Cepheids in
the Magellanic System. The sample consists of 338 objects: 285 and 53
variables in the Large and Small Magellanic Clouds, respectively. Based on
the pulsation periods and light-curve morphology, we classified 118 of our
type~II Cepheids as BL~Herculis, 120 as W~Virginis, 34 as peculiar
W~Virginis, and 66 as RV~Tauri stars. For all objects, we publish
time-series {\it VI} photometry obtained during the OGLE-IV survey, from
2010 to the end of 2017.

We present the most interesting individual objects in our collection:
16 type II Cepheids showing additional eclipsing or ellipsoidal
variability, two RV~Tau variables more than 2.5~mag fainter than other
stars of this type in the LMC, an RVb star that drastically decreased
the amplitude of the long-period modulation, type II Cepheids
exhibiting significant amplitude and period changes, and an RV~Tau
star which experiences interchanges of deep and shallow minima. We
show that peculiar W~Vir stars have markedly different spatial
distribution than other subclasses of type II Cepheids, which
indicates different evolutionary histories of these objects.}
{Stars: variables: Cepheids -- Stars: oscillations (including pulsations)
-- Stars: Population II -- Magellanic Clouds -- Catalogs}

\Section{Introduction}
Type II Cepheids are low-mass radially pulsating stars belonging to
the halo and old disk stellar populations. They are a relatively small
group of variable stars compared to other types of classical
pulsators: RR Lyrae stars or classical Cepheids, but they found
several important astrophysical applications. Type II Cepheids, like
other classical pulsators, obey a well-defined period--luminosity (PL)
relation, located 1.5--2 magnitudes below the PL relation followed by
classical Cepheids.

Type II Cepheids have been traditionally divided into three classes:
BL~Her, W~Vir, and RV~Tau stars. BL~Her stars, with the shortest
pulsation periods (1--4~d), evolve from the horizontal branch to the
asymptotic giant branch (AGB). W~Vir stars with intermediate periods
(4--20~d) are thought to experience thermal pulses which result in
blue loops into the instability strip region in the
Hertzsprung-Russell (HR) diagram, however a convincing evolutionary
model that would quantitatively describe this behavior still does not
exist (see discussion in Groenewegen and Jurkovic 2017a). RV~Tau
variables with the longest periods ($>20$~d) are post AGB stars, just
prior to the expulsion of planetary nebulae. They cross the
high-luminosity extension of the Cepheid instability strip. The
defining feature of RV~Tau stars are alternating deep and shallow
minima in their light curves, but usually they show strong
cycle-to-cycle variability. Closely related to the RV~Tau stars are
yellow semiregular variables (SRd stars) -- pulsating stars that
occupy the same area in the color--magnitude and PL diagrams, but do
not show clear alternations of minima. Sometimes this may be an effect
of strong cycle-to-cycle chaotic variations that mask the alternations
or quick interchanges of deep and shallow minima that also occur in
RV~Tau stars. In this paper, both subclasses of long-period type~II
Cepheids are referred to as RV~Tau stars.

In addition to these three groups, Soszyñski \etal (2008) established
another subtype of type~II Cepheids -- peculiar W~Vir stars. They
reveal distinct light curves and are usually brighter and bluer than
``regular'' W~Vir variables. About 50\% of the peculiar W~Vir stars in
the Magellanic Clouds show signs of binarity: eclipsing or ellipsoidal
modulation superimposed on the pulsation light curves or cyclic period
changes possibly caused by the light-time travel effect (Groenewegen
and Jurkovic 2017a). This suggests that peculiar W~Vir stars are
related to the so called binary evolution pulsators (BEPs) -- a class
of pulsating stars discovered in the OGLE Collection of Variable Stars
by Pietrzyñski \etal (2012). Recently, Karczmarek \etal (2017)
investigated possible contamination of BEPs among known samples of RR
Lyr stars and classical Cepheids and obtained values of 0.8\% and 5\%,
respectively. Since the number of type~II Cepheids is much smaller
than their classical siblings, one can suppose that the contamination
of BEPs among type~II Cepheids should be much larger than 5\%.

Soszyñski \etal (2008, 2010) published catalogs of type~II Cepheids in
the Magellanic Clouds obtained on the basis of the photometric data
collected by the OGLE-III project. The Large Magellanic Cloud (LMC)
sample consisted of 197 objects, later updated to 203 variables, while
in the Small Magellanic Cloud (SMC) we detected 43 such pulsators in
total. These type~II Cepheids, in particular their PL,
period--luminosity--color, and period--Wesenheit index relations, were
extensively studied (\eg Groenewegen and Jurkovic 2017b, Manick \etal
2017), also at near-infrared wavelengths (Matsunaga \etal 2009, 2011,
Ciechanowska \etal 2010, Ripepi \etal 2015, Bhardwaj \etal 2017), and
used for distance determinations inside our Galaxy and beyond
(Majaess \etal 2009ab, 2010, Ciechanowska \etal 2010). Smolec (2016)
compared his theoretical PL relation with the relation observed by
OGLE. Spectral energy distributions for the OGLE type~II Cepheids were
constructed and studied by Groenewegen and Jurkovic (2017a), who found
that about 60\% of the RV~Tau stars and about 10\% of the W~Vir stars
show an infrared excess. Manick \etal (2018) found that all RV~Tau
stars with a clear infrared excess have disc-type spectral energy
distribution, which suggests their binarity. The OGLE {\it I}-band
light curves were used by Kiss and Bódi (2017) to study the RVb
phenomenon (additional long-period variations of the mean brightness)
in some RV~Tau stars. They concluded that the long-term light
modulation can be fully explained by periodic obscurations by a dusty
disk around a binary system. The OGLE RV~Tau stars were also used as
tracers of the post-AGB stars (van Aarle \etal 2011, Woods \etal 2011,
Kamath \etal 2014).

In this paper, we extend the OGLE-III collection of type~II Cepheids
in the Magellanic Clouds by objects identified in the OGLE-IV
fields. We also supplement the OGLE-III light curves of the previously
known variables with the newest OGLE observations, increasing the
time-span of the OGLE photometry up to 21~years.

The paper is structured as follows. Section 2 provides a description
of the observations and data reduction pipeline. Section 3 presents
methods used for the identification and classification of type~II
Cepheids. In Section~4, we describe the collection and compare it with
other samples of type~II Cepheids in the Magellanic Clouds. In
Section~5, we discuss some interesting features of the published
samples of Cepheids and present the most interesting individual
objects in our collection. Section 6 summarizes our results.

\Section{Observations and Data Reduction}
OGLE observations of the Magellanic Clouds span more than two decades,
although this study is based on about eight years of the OGLE-IV
photometric monitoring: from March 2010 to December 2017. The light
curves collected during previous stages of the OGLE project can be
downloaded from the catalogs of Soszyñski \etal (2008, 2010)
and merged with the OGLE-IV data, however one should take into account
some possible offsets of the photometric zero points that can occur in
individual objects.

The OGLE-IV observations have been acquired using the 1.3-m Warsaw
telescope at Las Campanas Observatory, Chile, operated by the Carnegie
Institution for Science. The telescope is equipped with a mosaic
camera consisting of 32 CCD chips ($2048\times4096$~pixels) providing
a total field of view of 1.4 square degrees with a pixel scale of
0.26~arcsec. For stars located in technical gaps between the CCD
detectors, we used the light curves from the OGLE-IV auxiliary
photometric databases (Soszyñski \etal 2017a). Most of the
observations (typically 700 measurements) were taken with the standard
{\it I}-band filter from the Cousins photometric system, while the
remaining observations (typically 120 and up to 267 points) were
secured in the Johnson {\it V} passband.

The Difference Image Analysis package (Alard and Lupton 1998,
Wo¼niak 2000) was used to obtain the instrumental photometry of
all point sources. The light curves were then converted to
Johnson-Cousins {\it VI} magnitudes using the transformations given in
Udalski \etal (2015). We refer the reader to Udalski \etal (2015) for
a detailed description of the data reduction and calibration.

\Section{Selection and Classification of Type II Cepheids}
Type II Cepheids in the Magellanic Clouds were extracted from the
OGLE-IV photometric database using essentially the same methods that
were applied to other classical pulsators (\eg Soszyñski \etal
2015, 2017a). First, a period search for all {\it I}-band light curves
was performed through the Fourier analysis implemented in the {\sc
Fnpeaks}
code\footnote{http://helas.astro.uni.wroc.pl/deliverables.php?lang=en\&active=fnpeaks}.
Second, the light curves with the largest signal-to-noise ratios were
evaluated by visual inspection and initially divided into three
groups: pulsating stars, eclipsing binaries, and other variable
objects. Third, from the group of pulsating stars we distinguished
candidates for type~II Cepheids based on their positions in the PL and
color--magnitude diagrams, as well as characteristics of their light
curves quantified by the Fourier coefficients.

Type II Cepheids were subdivided into four separate classes: BL~Her,
W~Vir, peculiar W~Vir, and RV~Tau stars, depending on their pulsation
periods and light curve shapes. Stars with periods between 1~d and 4~d
were classified as BL~Her stars, between 4~d and 20~d as W~Vir and
peculiar W~Vir stars, and longer period variables were designated as
RV~Tau stars\footnote{For all Cepheids we provide ``single''
periods, \ie intervals between successive minima, also when the
period-doubling effect is present.}. We emphasize that periods of
different subclasses of type~II Cepheids partly overlap, so the
adopted boundary periods have just statistical meaning and for
individual objects one can use another classification. ``Regular'' and
peculiar W~Vir stars were distinguished based primarily on the
parameters of the light curve Fourier decomposition (Fig.~1), but we
also took into account mean brightness and colors (Fig.~2) of the
stars (peculiar W~Vir stars are on average brighter and bluer that
their ``regular'' siblings).

\begin{figure}[p]
\includegraphics[width=12.7cm]{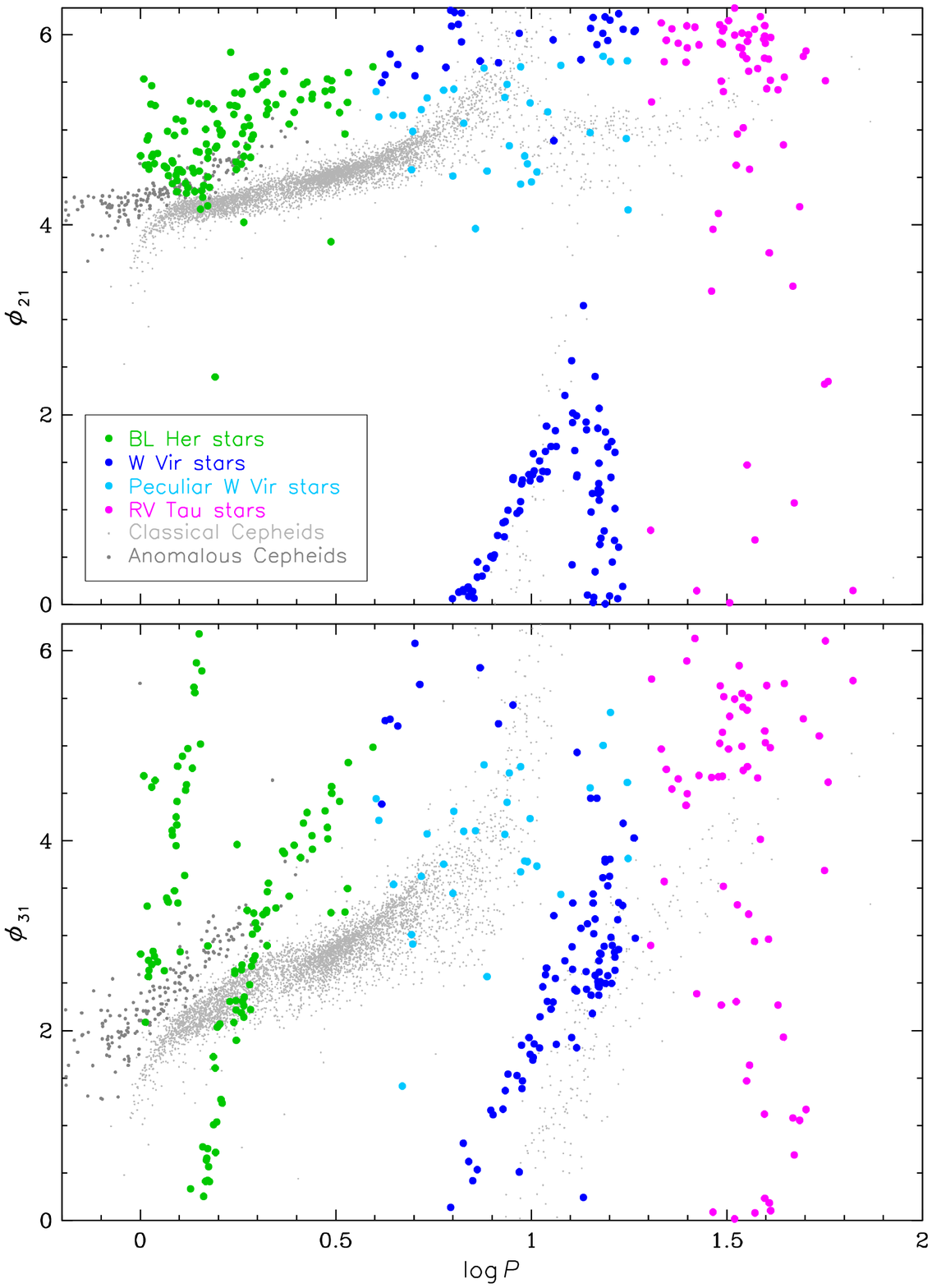}
\FigCap{Fourier coefficients $\phi_{21}$ and $\phi_{31}$ as a function of
the pulsation periods for Cepheids in the Magellanic Clouds. Green,
blue, light blue, and magenta points indicate BL~Her, W~Vir, peculiar
W~Vir, and RV~Tau stars, respectively. Large and small gray points,
respectively, show the location of classical and anomalous Cepheids
pulsating in the fundamental mode. Labels show two candidates for
ultra faint RV~Tau stars described in Section 5.4.}
\end{figure}

\begin{figure}[p]
\begin{center}
\includegraphics[width=12.7cm]{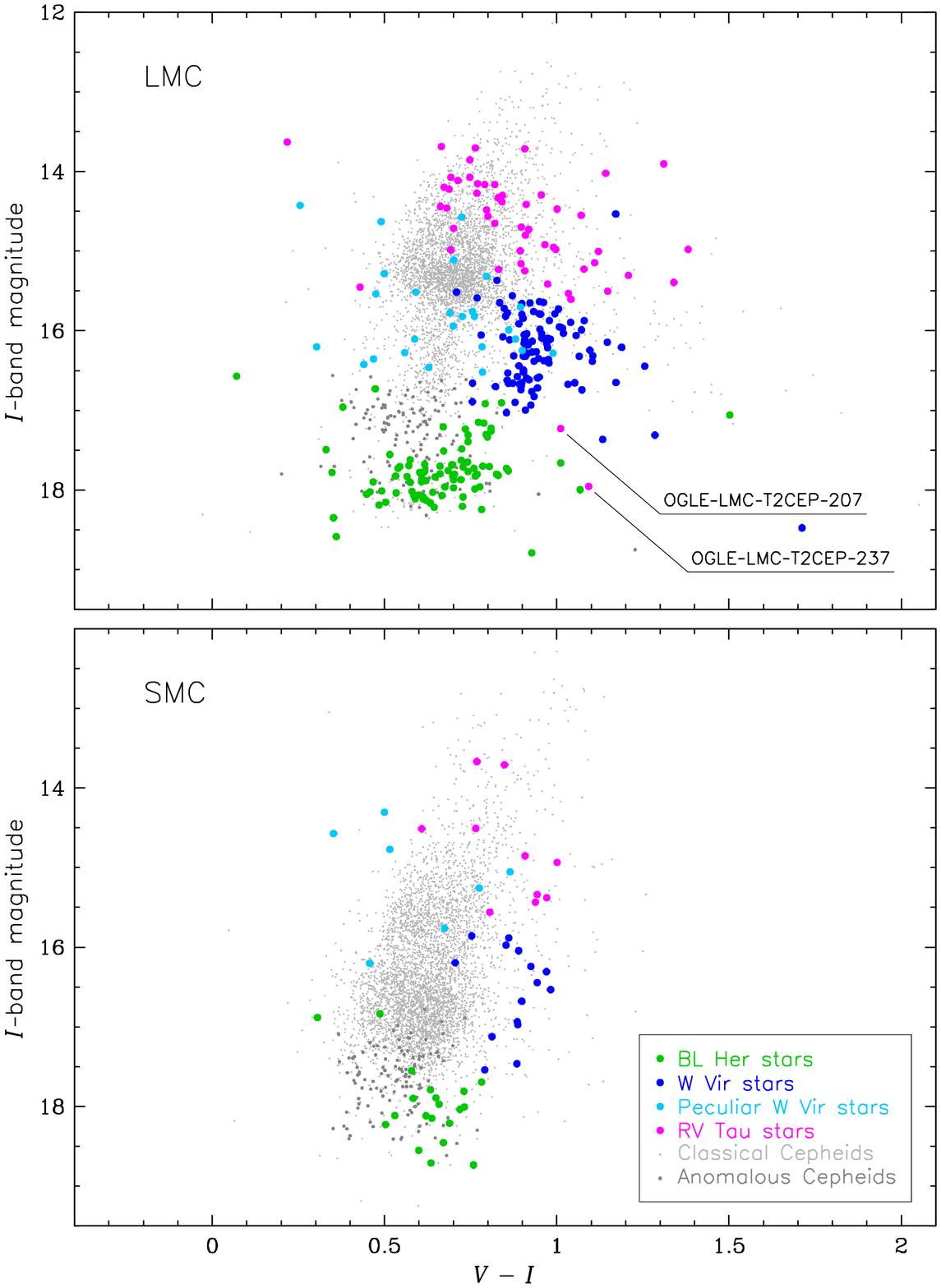}
\end{center}
\vspace*{-3mm}
\FigCap{Color $(V-I)$ vs. magnitude $I$ diagram for various types of
Cepheids in the LMC ({\it upper panel}) and SMC ({\it lower
panel}). Different colors correspond to the same types of pulsators as
in Fig.~1. Labels show two candidates for ultra faint RV~Tau stars
described in Section 5.4.}
\end{figure}

The new OGLE-IV photometry allowed us to verify the classification of
the candidates for type~II Cepheids published in the OGLE-III catalog
of variable stars (Soszyñski \etal 2008, 2010). Five objects
were reclassified as other types of variable stars and removed from
the OGLE collection of type~II Cepheids in the Magellanic System. In
Table 1 we list these stars with their new classification. Several
other candidates for type~II Cepheids were considered uncertain and
flagged in the remarks of our collection.

\MakeTableee{
l@{\hspace{12pt}} l@{\hspace{8pt}}}{12.5cm}
{Reclassified objects from the OGLE-III catalog of type~II Cepheids in
the Magellanic Clouds}
{\hline \noalign{\vskip3pt}
\multicolumn{1}{c}{Identifier} & \multicolumn{1}{c}{New classification} \\
\noalign{\vskip3pt}
\hline
\noalign{\vskip3pt}
OGLE-LMC-T2CEP-114 & Anomalous Cepheid    \\
OGLE-LMC-T2CEP-150 & Other                \\
OGLE-LMC-T2CEP-164 & Eclipsing binary     \\
OGLE-LMC-T2CEP-202 & Long-period variable \\
OGLE-SMC-T2CEP-21  & Other                \\
\noalign{\vskip3pt}
\hline}

\Section{The OGLE Collection of Type II Cepheids in the Magellanic System}

The OGLE collection of type~II Cepheids in the Magellanic System is
composed of 118 BL~Her, 120 W~Vir, 34 peculiar W~Vir, and 66 RV~Tau
variables -- in total 338 objects, of which 285 are located in the
LMC, and 53 in the SMC. Five of our type~II Cepheids (four BL~Her
stars and one W~Vir star) are probable members of four LMC globular
clusters (NGC~1786, NGC~1835 -- contains two Cepheids, NGC~1939, and
Hodge 11), as the sky positions of these objects coincide with the
positions of the clusters.

The entire collection can be downloaded {\it via} the WWW interface or
from the FTP site:
\begin{center}
{\it http://ogle.astrouw.edu.pl}\\
{\it ftp://ftp.astrouw.edu.pl/ogle/ogle4/OCVS/lmc/t2cep/}\\
{\it ftp://ftp.astrouw.edu.pl/ogle/ogle4/OCVS/smc/t2cep/}\\
\end{center}

The identifiers of our type~II Cepheids follow the scheme introduced
in the OGLE-III catalog (OGLE-LMC-T2CEP-NNN or OGLE-SMC-T2CEP-NN). The
newly detected variables have numbers larger than 203 and 43 in the
LMC and SMC, respectively, and are organized by increasing right
ascension. For each star, we provide, among others, its equatorial
coordinates (J2000.0), the cross-matches with the extragalactic part
of the General Catalogue of Variable Stars (GCVS, Artyukhina \etal
1995), mean magnitudes in the {\it I}- and {\it V}-bands, pulsation
period, epoch of the maximum light, peak-to-peak {\it I}-band
amplitude, and Fourier parameters derived from the {\it I}-band light
curves. The pulsation periods have been adjusted with the {\sc Tatry}
code of Schwarzenberg-Czerny (1996), based solely on the photometry
collected in 2010-2017, from the beginning of the fourth phase of the
OGLE project. For some type~II Cepheids, in particular W~Vir and
RV~Tau stars, periods derived from light curves collected by the
OGLE-II (1997-2000) or OGLE-III (2001-2009) surveys may be different
than provided in this collection, because large and erratic period
changes is a common feature in these objects (see Section~5.4). To
obtain the amplitudes and intensity-averaged magnitudes of the
variables, we fitted the light curves with a cosine Fourier series.

\begin{figure}[p]
\begin{center}
\includegraphics[width=12.7cm]{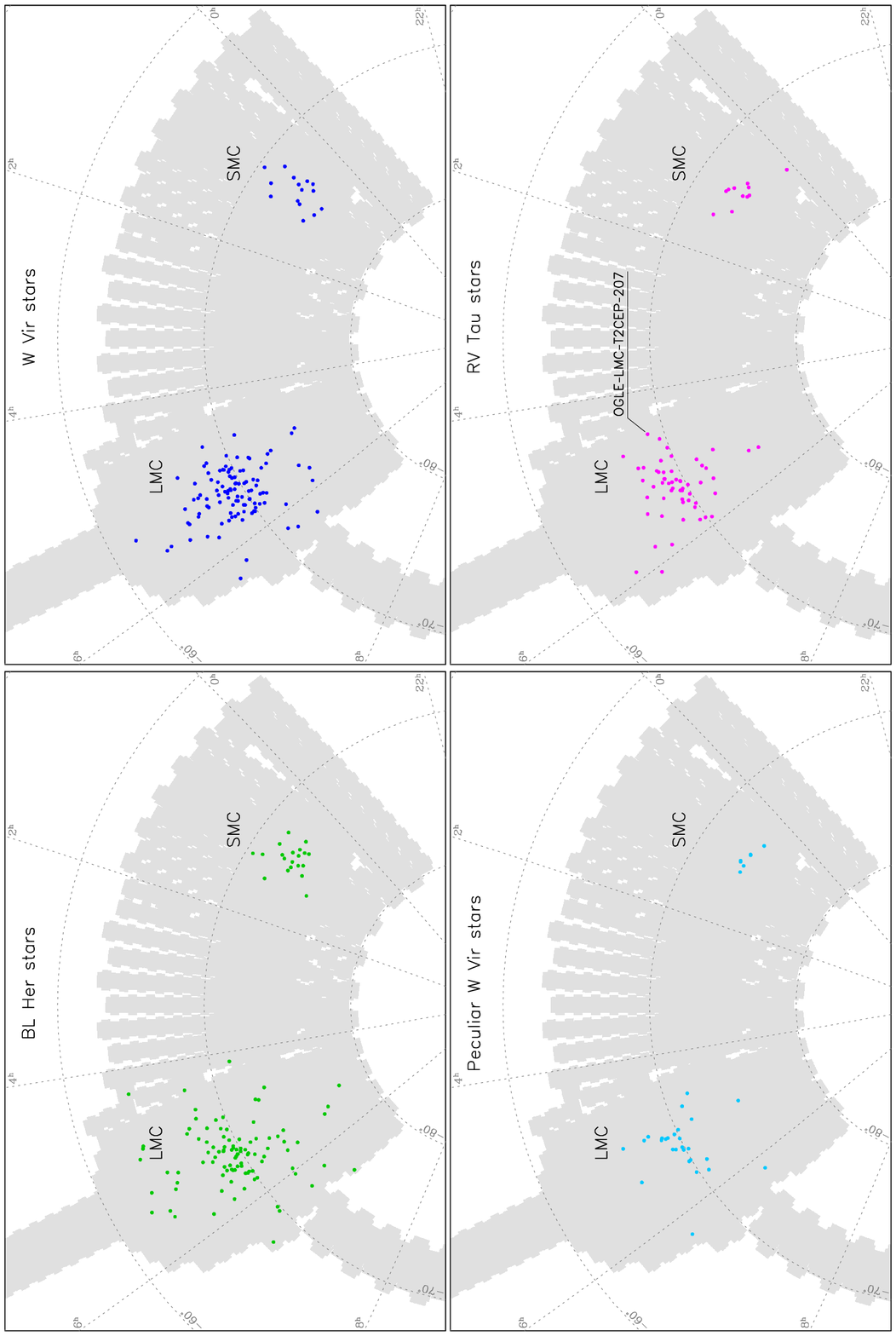}
\end{center}
\vspace*{-3mm}
\FigCap{Sky distribution of type~II Cepheids in the Magellanic
Clouds. Each panel shows positions of different subclasses of type~II
Cepheids. The gray area shows the OGLE-IV footprint.}
\end{figure}

Sky maps showing the positions of the four classes of type~II Cepheids
are presented in Fig.~3. At the first glance, BL~Her stars show the
largest dispersion of their positions around the centers of the LMC
and SMC, W~Vir stars seem to be slightly more concentrated toward the
centers, RV~Tau stars are even more focused around the central parts
of both galaxies, and the peculiar W~Vir stars populate mostly the bar
of the LMC and the very center of the SMC. The spatial distribution
would be the key to understand of the origin and current evolutionary
status of various classes of type~II Cepheids. It is worth
noting that type~II Cepheids avoid the region between both Clouds --
the so called Magellanic Bridge. In this regard, type~II Cepheids
differ from anomalous Cepheids which present very broad halo.

\Subsection{Cross-Match with Other Catalogs}
In order to test the completeness of our collection of type~II
Cepheids in the Magellanic Clouds, we cross-matched it with the GCVS
(Artyukhina \etal 1995), the catalog of periodic variable stars
detected from the EROS-2 survey (Kim \etal 2014), and the catalog of
Cepheids recently published as a part of the {\it Gaia} Data Release 2
(DR2, Clementini \etal 2018, Holl \etal 2018).

The extragalactic part of the GCVS (Artyukhina \etal 1995) contains 28
stars classified as type~II Cepheids (of type CWA, CWB, RV, RVA) in
the LMC and SMC. A preliminary version of the OGLE collection
contained 21 of these objects. We carefully inspected the OGLE light
curves of the remaining seven sources and decided to supplement our
collection with one of these objects (HV~2522 = OGLE-LMC-T2CEP-251),
which seems to be a yellow semiregular variable (SRd star). Other
candidates for type~II Cepheids from the GCVS are classical Cepheids,
long-period variables or eclipsing binaries.

Kim \etal (2014) used automatic algorithms to select 117\,234
candidates for periodic variable stars from the EROS-2 LMC photometric
database. This list includes 343 objects classified as type~II
Cepheids, of which 182 stars were cataloged in the OGLE
Collection. Most of the remaining candidates have their counterparts
in the OGLE photometric database within 1~arcsec search radius, but
only one additional object (lm0336k4219 = OGLE-LMC-T2CEP-246) was
recognized as a real type~II Cepheid (a W~Vir star with nearly
symmetrical light curve) and was included to our sample. The remaining
EROS-2 candidates for type~II Cepheids represent a wide spectrum of
stellar variability classes: anomalous Cepheids, eclipsing binaries,
long-period variables, spotted stars, etc.

A cross-match of our sample with the Gaia DR2 catalog of 191 potential
type~II Cepheids in the Magellanic Clouds\footnote{We defined the
region of the Magellanic Clouds as a box $0^{\circ}<R.A.<100^{\circ}$,
$-80^{\circ}<Dec<-60^{\circ}$ -- a larger area than that defined by
Clementini \etal (2018).} (Clementini \etal 2018) did not bring new
discoveries. This and the previously published parts of the OGLE
Collection of Variable Stars contain in total 159 objects from the
Gaia list: 138 true type~II Cepheids, 16 classical Cepheids, and five
anomalous Cepheids (Soszyñski \etal 2017a). We successfully
identified OGLE-IV light curves for all 32 remaining Gaia candidates
for type~II Cepheids and, after a visual inspection, we classified
them as long-period variables, eclipsing or ellipsoidal binaries,
irregular variables or other types of variable stars.

These tests show that the completeness of our collection of type~II
Cepheids in the Magellanic Clouds is very high, close to 100\%. We
suppose that at most several variables may be located outside the
OGLE-IV fields, in the far outskirts of the Magellanic System. Also,
it cannot be excluded that a few type~II Cepheids were overlooked due
to the unusual morphology of their light curves or the location in the
immediate vicinity of bright stars.

\begin{figure}[p]
\begin{center}
\includegraphics[width=12.7cm]{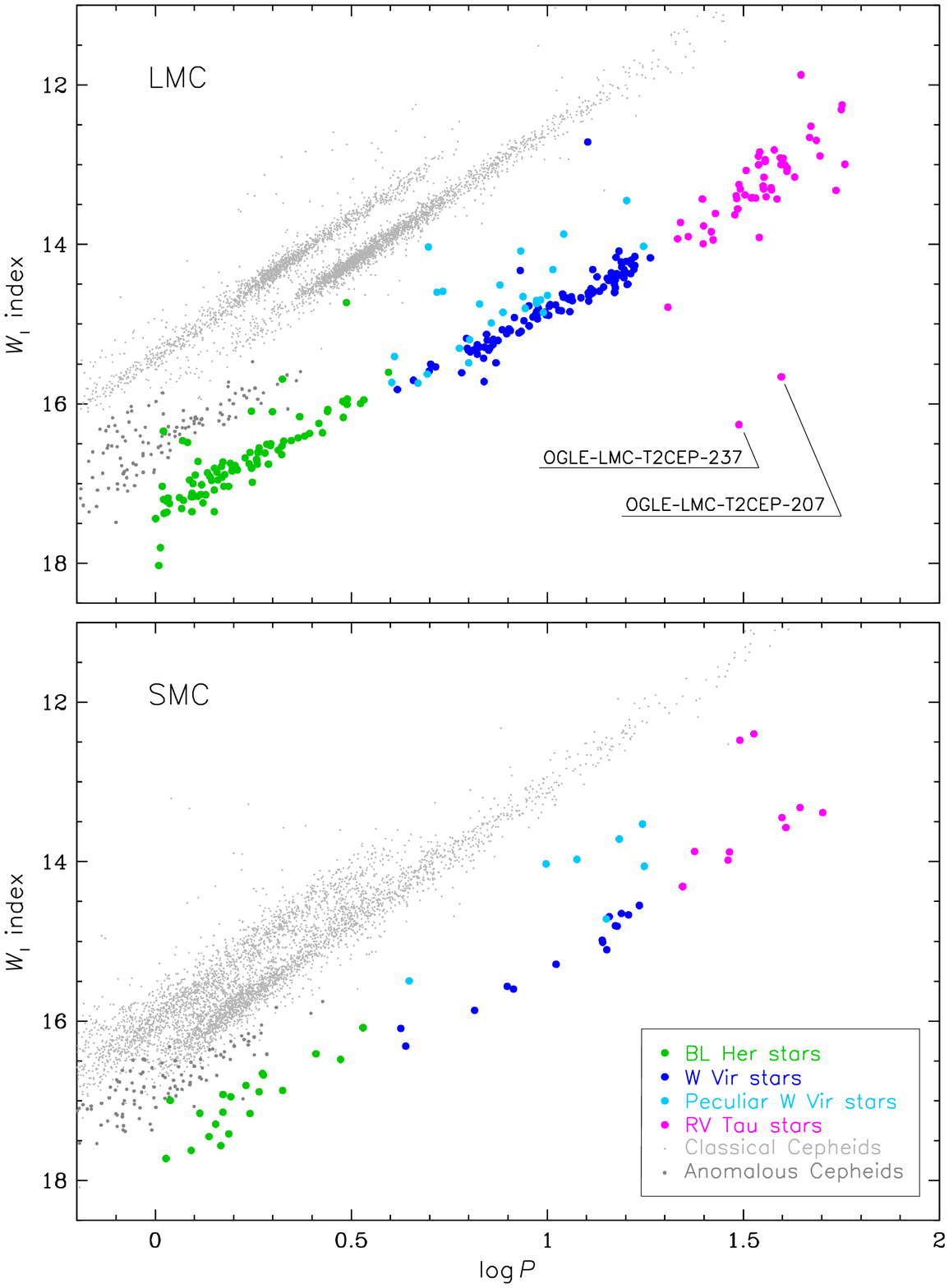}
\end{center}
\vspace*{-3mm}
\FigCap{Period vs. Wesenheit index diagram for various types of
Cepheids in the LMC ({\it upper panel}) and SMC ({\it lower
panel}). Different colors correspond to the same types of pulsators as
in Fig.~1.}
\end{figure}

\Section{Discussion}

\Subsection{Period--Luminosity Relation}
The PL relation obeyed by type~II Cepheids is probably the
most-studied feature of these stars in the Magellanic Clouds. Fig.~4
presents PL diagrams for type~II Cepheids in the LMC (upper panel) and
SMC (lower panel), where as a ``luminosity'' we used a
reddening-independent Wesenheit index, $W_I=I-1.55(V-I)$. For
comparison, classical and anomalous Cepheids (Soszyñski \etal
2015, 2017a) are also plotted in Fig.~4.

Our nearly complete PL diagrams confirm earlier findings. In the LMC,
BL~Her and W~Vir stars are roughly co-linear in the period--$W_I$
plane, while RV~Tau stars are located above the regression line fitted
to the BL~Her and W~Vir variables and show larger scatter of the PL
sequence. Manick \etal (2016) attributed this to the extinction by
circumstellar dust. On the other hand, in the SMC, BL Her stars seem
to be located above the PL relation delineated by ``regular'' W~Vir
and RV~Tau stars. Peculiar W~Vir stars are on average brighter than
``regular'' W~Vir stars, which can be partly attributed to the
additional light from their potential companions (see Section~5.2),
although note that the solution of Pilecki \etal (2017) for
OGLE-LMC-T2CEP-098 (a peculiar W~Vir star in an eclipsing binary
system) placed this pulsating star about 1~mag above the PL relation
for type~II Cepheids.

A number of outliers visible among ``ordinary'' BL~Her, W~Vir, and
RV~Tau variables may also be explained by the blending by unresolved
stars, although some of these stars are uncertain objects, which is
flagged in the remarks of the collection. Two faint RV~Tau stars
labeled in the upper panel of Fig.~4 are described in Section~5.4.

\begin{figure}[p]
\begin{center}
\includegraphics[width=12.7cm]{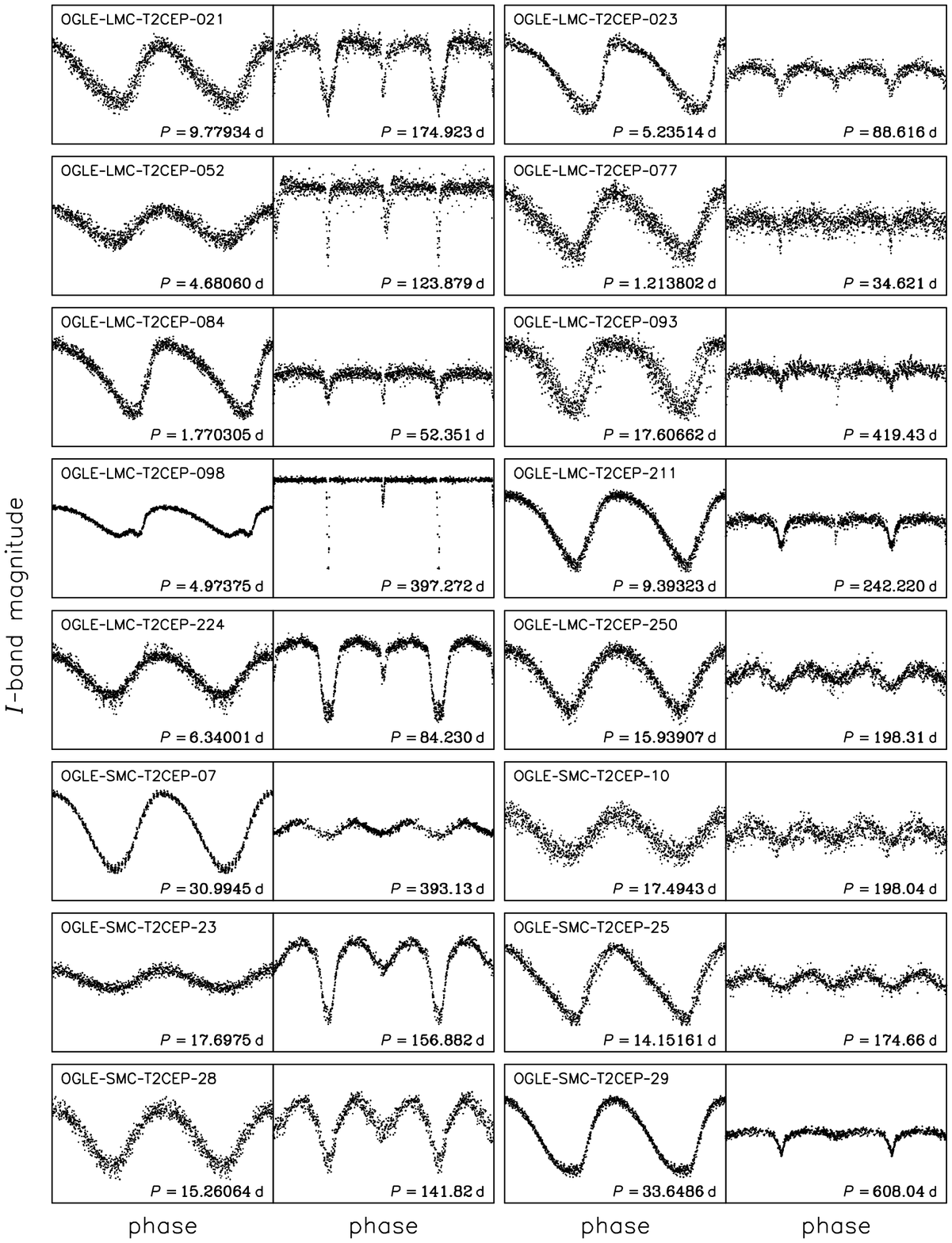}
\end{center}
\vspace*{-3mm}
\FigCap{Disentangled {\it I}-band light curves of type~II Cepheids
with additional eclipsing or ellipsoidal variability. Each object is
presented in two panels showing pulsating light curves ({\it left
panels}) and eclipsing or ellipsoidal light curves ({\it right
panels}). All the data come from the OGLE-IV database with the
exception of OGLE-LMC-T2CEP-077 which light curve was taken from the
OGLE-III database (there are very few OGLE-IV observations for this
star).}
\end{figure}

\Subsection{Type II Cepheids in Binary Systems}
The OGLE-III catalog of type~II Cepheids in the Magellanic Clouds
(Soszyñski \etal 2008, 2010) contains as many as 13 pulsating
stars with eclipsing or ellipsoidal modulation. In the present
investigation, we supplement this group with three additional objects:
OGLE-LMC-T2CEP-211, OGLE-LMC-T2CEP-224, and OGLE-LMC-T2CEP-250. The
disentangled light curves of all 16 Cepheids with eclipsing or
ellipsoidal variations are presented in Fig.~5 and their periods and
identifiers (with additional identifiers from the OGLE collection of
eclipsing binary systems in the Magellanic Clouds, Pawlak \etal 2016)
are summarized in Table~2.

\MakeTableee{
l@{\hspace{8pt}} c@{\hspace{8pt}} c@{\hspace{8pt}} l@{\hspace{8pt}} c@{\hspace{8pt}}}{12.5cm}
{Type II Cepheids in eclipsing and ellipsoidal binary systems}
{\hline \noalign{\vskip3pt}
\multicolumn{1}{c}{Identifier} & Classification & Pulsation & \multicolumn{1}{c}{Identifier} & \multicolumn{1}{c}{Orbital} \\
\multicolumn{1}{c}{as a type~II Cepheid} &  & period [d] & \multicolumn{1}{c}{as a binary system} & \multicolumn{1}{c}{period [d]} \\
\noalign{\vskip3pt}
\hline
\noalign{\vskip3pt}
OGLE-LMC-T2CEP-021 & pec. W~Vir & 9.77937 & OGLE-LMC-ECL-28274 & 174.923 \\
OGLE-LMC-T2CEP-023 & pec. W~Vir & 5.23514 & OGLE-LMC-ECL-28388 & 88.616 \\
OGLE-LMC-T2CEP-052 & pec. W~Vir & 4.68061 & OGLE-LMC-ECL-30076 & 123.879 \\
OGLE-LMC-T2CEP-077 & BL~Her     & 1.21380 & OGLE-LMC-ECL-31119 & 34.621 \\
OGLE-LMC-T2CEP-084 & BL~Her     & 1.77031 & OGLE-LMC-ECL-31272 & 52.351 \\
OGLE-LMC-T2CEP-093 & pec. W~Vir & 17.6067 & OGLE-LMC-ECL-31746 & 419.43 \\
OGLE-LMC-T2CEP-098 & pec. W~Vir & 4.97375 & OGLE-LMC-ECL-31922 & 397.272 \\
OGLE-LMC-T2CEP-211 & pec. W~Vir & 9.39323 & OGLE-LMC-ECL-26653 & 242.220 \\
OGLE-LMC-T2CEP-224 & pec. W~Vir & 6.34001 & OGLE-LMC-ECL-29309 & 84.230 \\
OGLE-LMC-T2CEP-250 & pec. W~Vir & 15.9391 &                    & 198.31 \\
OGLE-SMC-T2CEP-07  & RV~Tau     & 30.9945 &                    & 393.13 \\
OGLE-SMC-T2CEP-10  & pec. W~Vir & 17.4943 &                    & 198.04 \\
OGLE-SMC-T2CEP-23  & pec. W~Vir & 17.6975 & OGLE-SMC-ECL-3123  & 156.882 \\
OGLE-SMC-T2CEP-25  & pec. W~Vir & 14.1516 &                    & 174.66 \\
OGLE-SMC-T2CEP-28  & pec. W~Vir & 15.2606 & OGLE-SMC-ECL-7243  & 141.82 \\
OGLE-SMC-T2CEP-29  & RV~Tau     & 33.6492 & OGLE-SMC-ECL-7250  & 608.04 \\
\noalign{\vskip3pt}
\hline}

In most of these objects, we are confident that the pulsating stars
belong to the binary systems (so they are not physically unrelated
blends observed by chance along the same line of sight), because we
can detect combination frequencies in their light curves. Usually, the
combination frequencies are equal to $(1/P_\mathrm{puls} \pm
1/P_\mathrm{orb})$ or $(1/P_\mathrm{puls} \pm 2/P_\mathrm{orb})$,
which reflects complex oscillations of stars which are distorted by
tidal interactions from their companions.

The majority of our eclipsing and ellipsoidal type~II Cepheids are
classified as peculiar W~Vir stars. The only exceptions are
OGLE-LMC-T2CEP-077 and OGLE-LMC-T2CEP-084, classified as a BL~Her
star, and OGLE-SMC-T2CEP-07 and OGLE-SMC-T2CEP-29, classified as
RV~Tau stars, but they are likely equivalents of the peculiar W~Vir
stars in the range of pulsation periods assigned to RV~Tau
variables. It is worth noting that Groenewegen and Jurkovic (2017a)
derived and investigated the observed minus calculated (O$-$C)
diagrams for the OGLE-III light curves of type~II Cepheids in the
Magellanic Clouds and found about 20 additional candidates for binary
Cepheids.

Careful spectroscopic analysis of the peculiar W~Vir stars in binary
systems may solve the problem of their origin. Recently, the
``Araucaria'' collaboration (Pilecki \etal 2017) presented a study of
OGLE-LMC-T2CEP-098 -- a peculiar W~Vir star in an eclipsing binary
system. The mass of the pulsating component was estimated to be
$1.51\pm0.09$~M$_\odot$ -- significantly larger than expected for
ordinary type~II Cepheids. It was suggested that the current
evolutionary status of the Cepheid and its companion could be
explained by episodes of mass transfer between the components of the
binary systems, and thus OGLE-LMC-T2CEP-098 meets the definition of
binary evolution pulsators (Pietrzyñski \etal 2012).

\Subsection{Peculiar W~Vir Stars}
Our collection contains 34 peculiar W~Vir stars in the Magellanic
Clouds, of which 35\% exhibit eclipsing or ellipsoidal
variations. Additional candidates for binaries among OGLE peculiar
W~Vir variables were found by Groenewegen and Jurkovic (2017a) through
the light-time effect, so the percentage of peculiar W~Vir stars with
signs of binarity reaches 50\%. This clearly indicates that binarity
is evolutionary essential to explain pulsations of peculiar W~Vir
stars.

Our distinction between ``regular'' and peculiar W~Vir stars was based
on their light curve shapes. In particular, both groups occupy
different regions in the period--$\phi_{21}$ and period--$\phi_{31}$
diagrams (Fig.~1). The same feature was recently used to distinguish
30 candidates for peculiar W~Vir stars in the Galactic bulge
(Soszyñski \etal 2017b). It is worth noting, however, that
compared to the Magellanic Clouds sample, the peculiar W~Vir stars in
the center of the Milky Way have a much smaller rate of eclipsing
binary systems. We found there only one type~II Cepheid
(OGLE-BLG-T2CEP-674) with the eclipsing modulation.

Matsunaga \etal (2009) noticed that peculiar W~Vir stars are absent in
globular clusters which suggests that this class of pulsating stars
may belong to a younger stellar population than other type~II
Cepheids. This conclusion is supported by the spatial distribution of
these stars in the LMC and SMC (Fig.~3). The vast majority of peculiar
W~Vir stars are located in the central regions of both galaxies, in
agreement with the distribution of a young stellar population (for
example classical Cepheids, Soszyñski \etal 2017a).

\begin{figure}[t]
\begin{center}
\includegraphics[width=10.0cm]{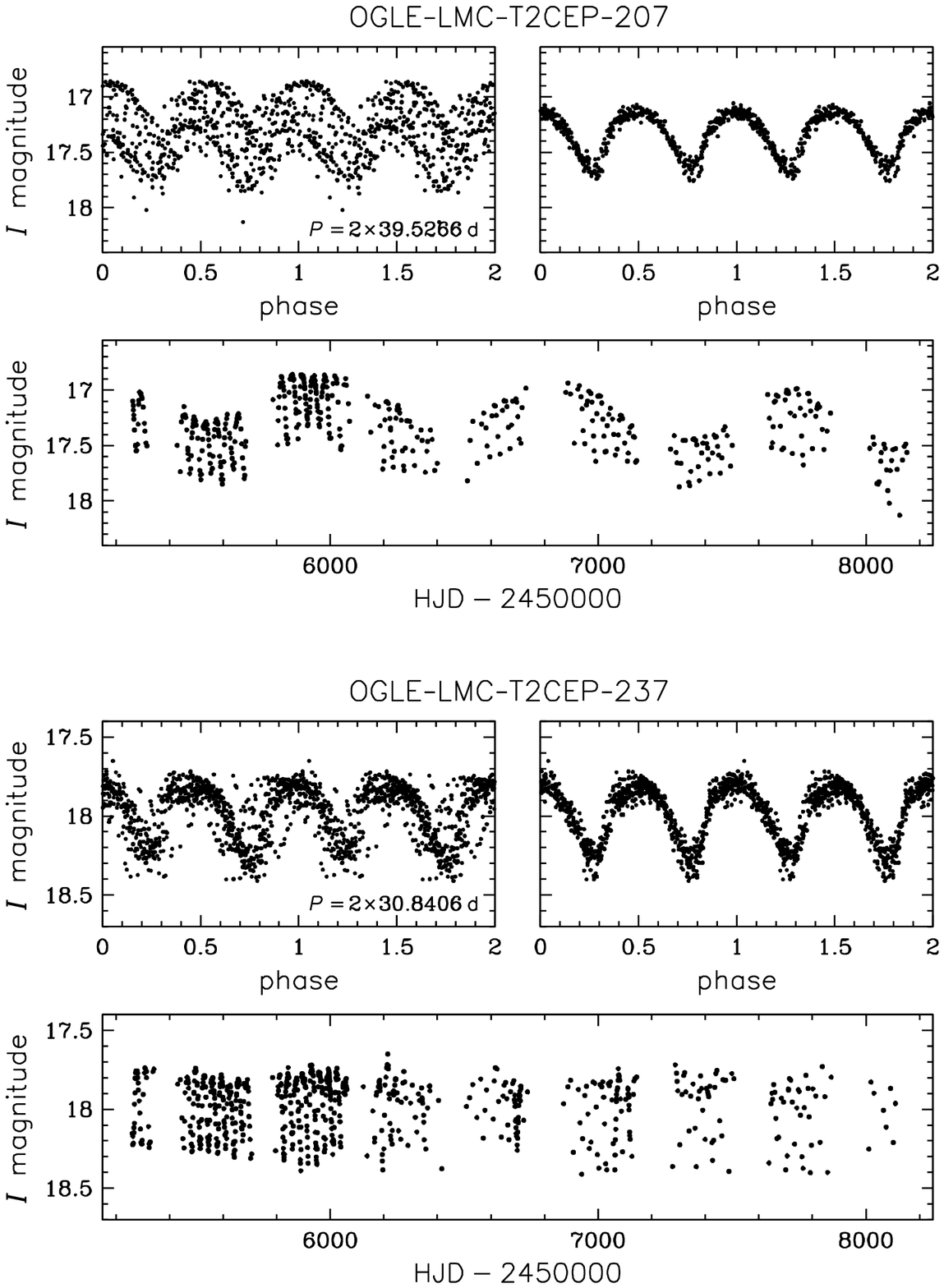}
\end{center}
\FigCap{Light curves of OGLE-LMC-T2CEP-207 and OGLE-LMC-T2CEP-237 --
candidates for extremely faint RV~Tau stars. For each star we plot the
original {\it I}-band light curve folded with the mean period ({\it
upper left panels}), the folded light curve corrected for the period
changes and the long-term brightness modulations ({\it upper right
panels}), and the unfolded light curve ({\it lower panels}).}
\end{figure}

\Subsection{Particularly Interesting Type II Cepheids}
Two candidates for RV~Tau stars in the LMC -- OGLE-LMC-T2CEP-207 and
OGLE-LMC-T2CEP-237 -- are surprisingly faint, more than 2.5~mag
fainter than other RV~Tau variables with the same periods (Fig.~4),
but with $(V-I)$ color indices typical for RV~Tau stars
(Fig.~2). Light curves of these two objects are shown in
Fig.~6. OGLE-LMC-T2CEP-207 experiences long-period (850~d) variations
of the mean brightness, which place this object in the group of RVb
stars. Both variables exhibit changes of periods, which is also a
common feature of RV~Tau stars. In the upper right panels of Fig.~6,
we present light curves of both stars corrected for the period changes
and the long-term modulation. In both stars, the alternations of
deeper and shallower minima are small but detectable, at least when
studying observations obtained in individual observing seasons.

If these objects are indeed RV~Tau stars, there are two explanations
of their origin. First, they are located in the LMC and belong to a
new subclass of ultra faint RV-Tau-like variables. Second, they are
classical RV~Tau stars, but located behind the LMC, more than three
times farther than the center of this galaxy. The positions of these
two stars is the sky (Fig.~3) suggest that they belong to the LMC,
although OGLE-LMC-T2CEP-207 is located outside the center of the LMC,
the most west of all RV~Tau stars in this galaxy. On the other hand,
OGLE-LMC-T2CEP-237 (not labeled in Fig.~3) is located in the sky at
the very center of the LMC. Further research is necessary to explain
the nature of these stars.

\begin{figure}[h]
\begin{center}
\includegraphics[width=12.7cm]{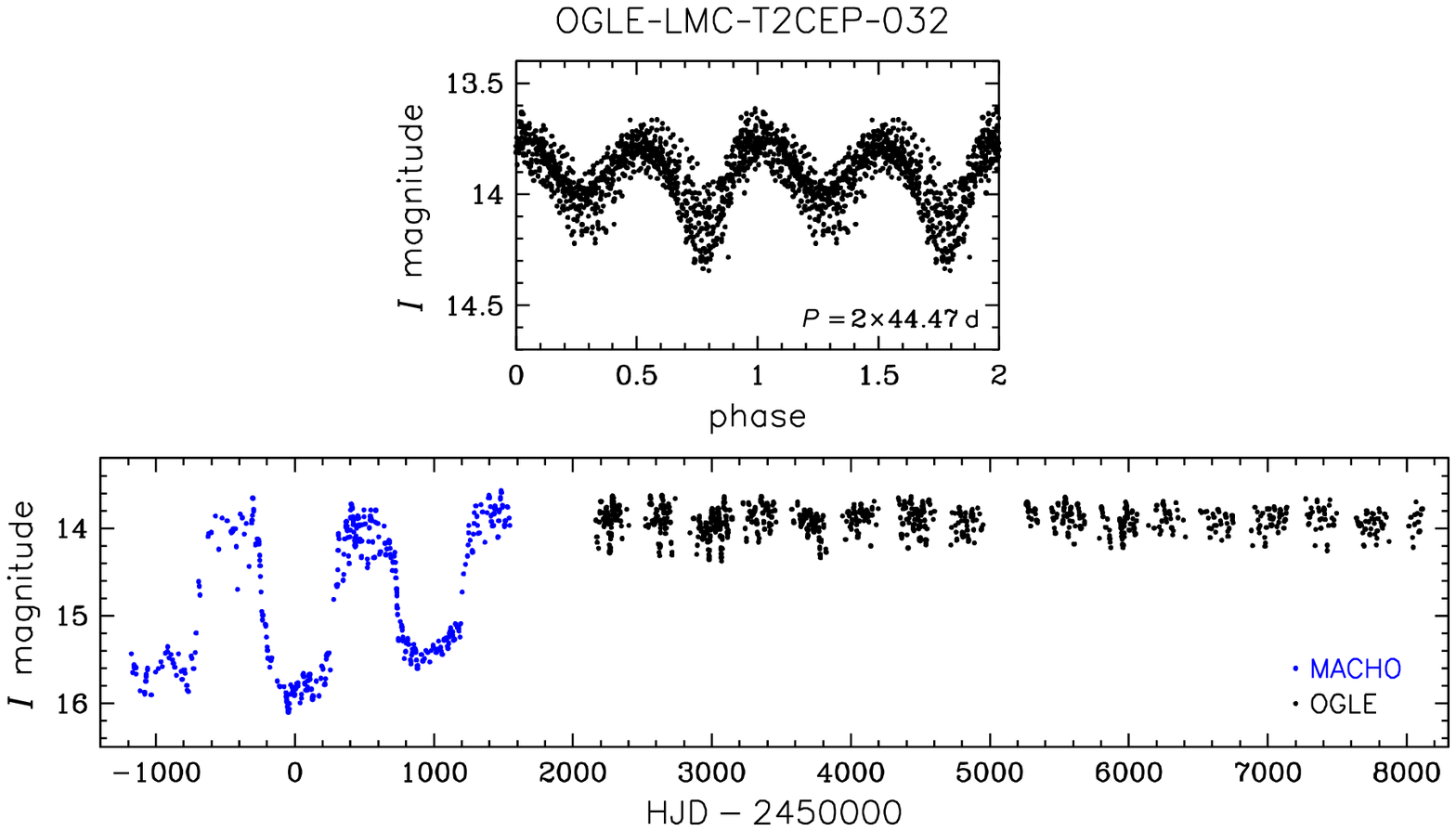}
\end{center}
\FigCap{MACHO and OGLE light curves of OGLE-LMC-T2CEP-032 -- an RVb
star which dramatically decreased the amplitude of the long-period
variations of the mean brightness. {\it Upper panel}: folded OGLE-III
and OGLE-IV {\it I}-band light curve. {\it Lower panel}: unfolded
light curve -- blue points indicate the MACHO $R_{\textrm{M}}$-band
magnitudes, black points indicate the OGLE light curve. The zero point
of the MACHO photometry was shifted to agree with the OGLE
magnitudes.}
\end{figure}

Beside OGLE-LMC-T2CEP-207, our collection contains only one RVb star
with large-amplitude modulations of the mean brightness --
OGLE-LMC-T2CEP-200. Additionally, a few RV~Tau stars show long-term
modulations of the mean luminosity, but with small amplitudes. One of
the most interesting objects of this type is OGLE-LMC-T2CEP-032, which
in 1990s was an RVb star with large (but variable) amplitudes of the
long-period (about 960~d) variability. Since at that time this star
was not monitored by the OGLE project, in Fig.~7 we plot its
$R_{\textrm{M}}$-band magnitudes obtained by the MACHO survey
(Alcock \etal 1993). Since 2001, when OGLE began regular observations
of this star, the long-period modulation is still detectable in the
light curve, but has a much smaller amplitude of about 0.1~mag.

The OGLE light curves of point sources in the Magellanic Clouds span
from 8 years (OGLE-IV) to 21 years (OGLE-II + OGLE-III + OGLE-IV),
depending on the region. These continuous, long-term observations are
an excellent tool for studying non-stationary processes in variable
stars, like period changes, amplitude modulations, or additional
periodicities.

\begin{figure}[t]
\begin{center}
\includegraphics[width=12.7cm]{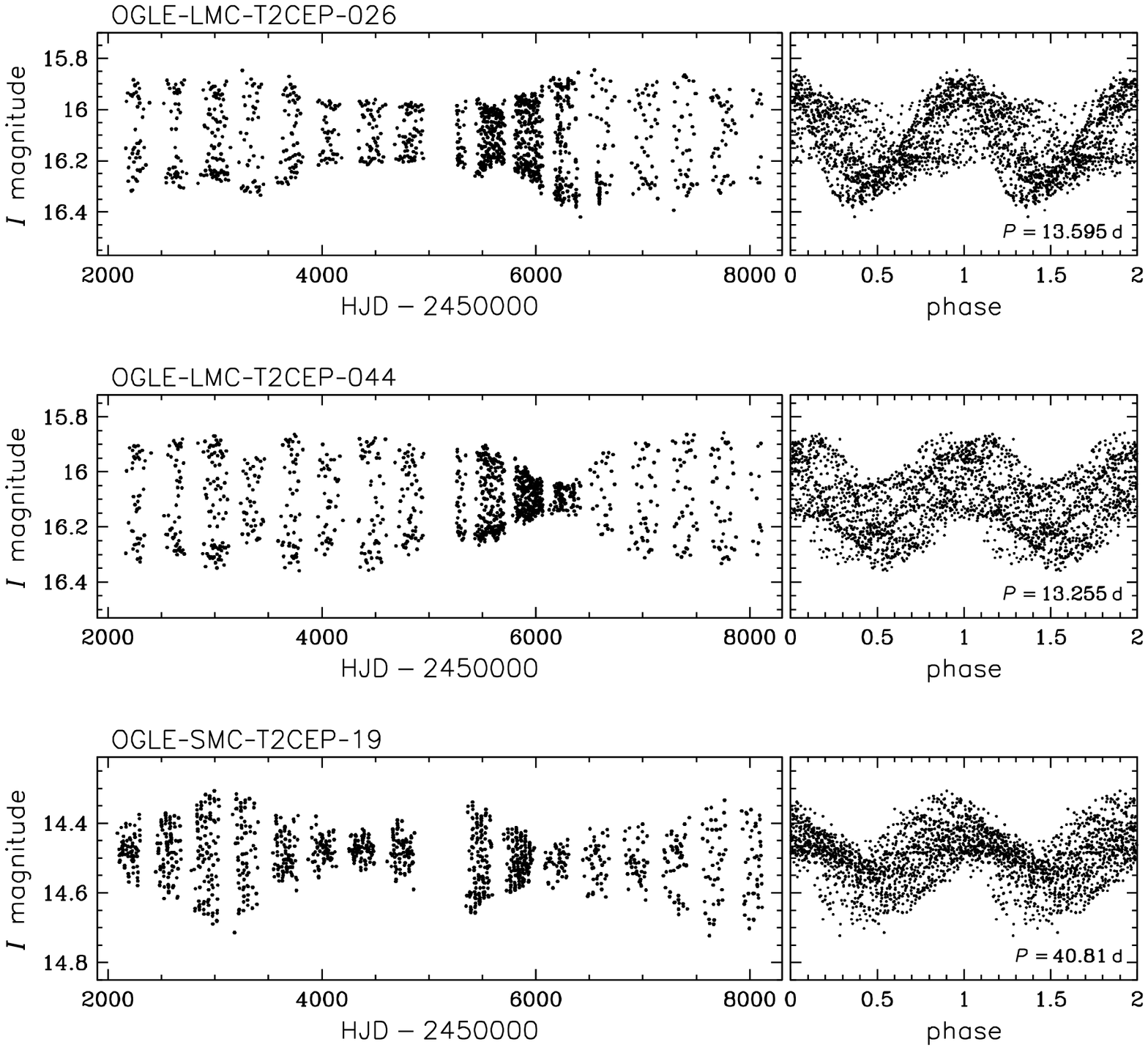}
\end{center}
\FigCap{Light curves of OGLE-LMC-T2CEP-026, OGLE-LMC-T2CEP-044, and
OGLE-SMC-T2CEP-19 -- type~II Cepheids with significant amplitude
changes. {\it Left panels} show unfolded OGLE-III and OGLE-IV {\it
I}-band light curves, {\it right panels} present the same light curves
folded with the mean pulsation periods. Note that the scatter of the
points is partly caused by the period changes experienced by all these
objects.}
\end{figure}

Unstable light curves are the most common among W~Vir and RV~Tau
stars, which is particularly evident when the OGLE-IV photometry is
merged with the older OGLE observations. Fig.~8 shows the OGLE-III +
OGLE-IV light curves of three type~II Cepheids with a significant
amplitude modulation. Note that in some cases the amplitude changes
are quasi-periodic (lower panel of Fig.~8), which resembles the
Blazhko effect in RR Lyr-type stars. Right panels of Fig.~8 present
the same light curves folded with the best matching constant
periods. The substantial scattering of points is caused not only by
the amplitude modulation, but also by the erratic period changes --
another very common feature of long-period type~II Cepheids.

\begin{figure}[p]
\begin{center}
\includegraphics[width=12.7cm]{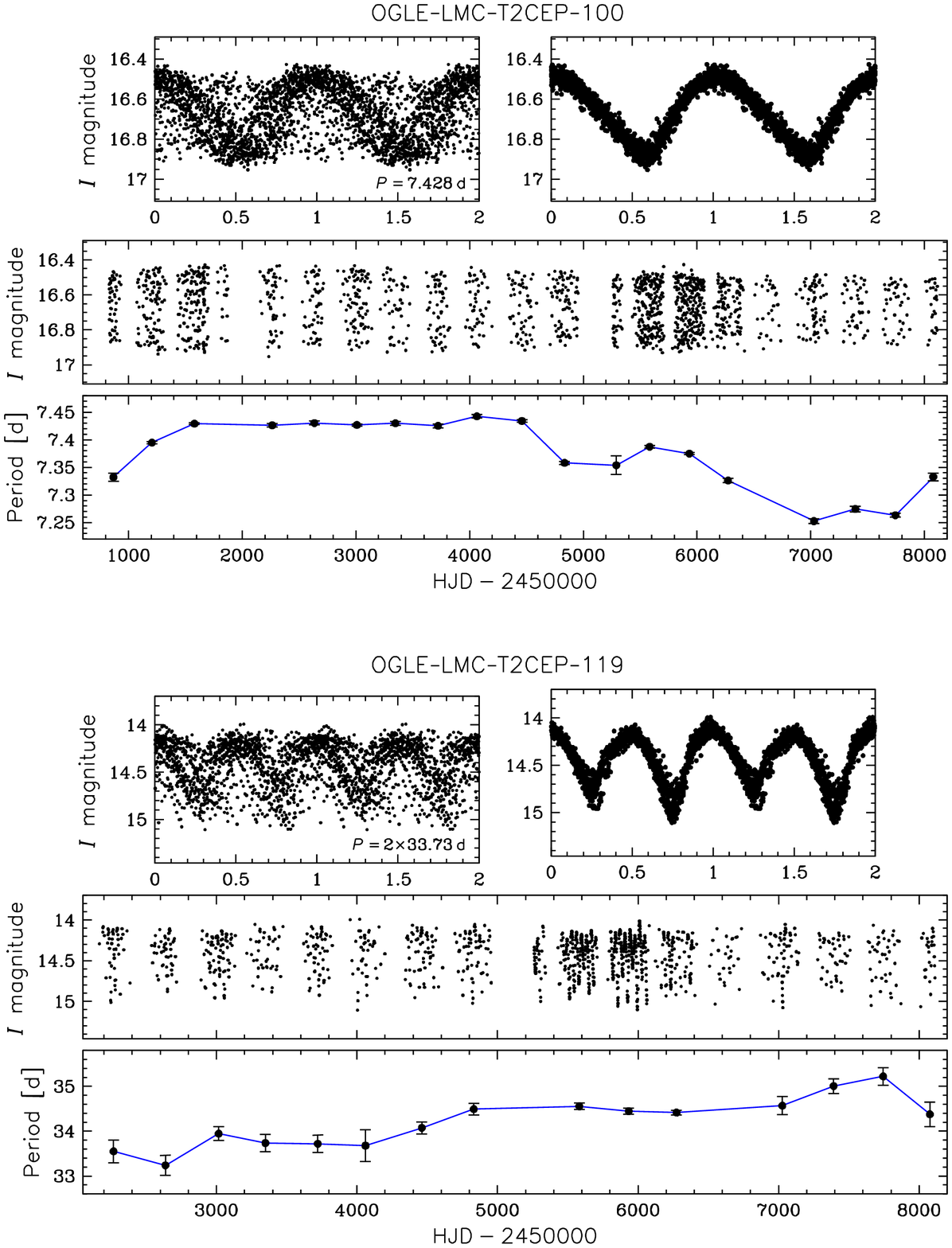}
\end{center}
\FigCap{Light curves of OGLE-LMC-T2CEP-100 and OGLE-LMC-T2CEP-119 --
type~II Cepheids with significant period changes. For each star we
plot the original {\it I}-band light curve folded with the mean period
({\it upper left panels}), the folded light curve corrected for the
period changes ({\it upper right panels}), the unfolded light curve
({\it middle panels}), and a diagram showing changes of the period
with time ({\it lower panels}).}
\end{figure}

In Fig.~9, we present other two examples of period-changing variables:
a W~Vir star OGLE-LMC-T2CEP-100 and an RV~Tau star
OGLE-LMC-T2CEP-119. In both cases, the light curves cannot be properly
phased using a constant period (upper left panels), but phasing with
periods measured separately in each observing season produces well
defined light curves (upper right panels). Period changes in W~Vir and
RV~Tau stars are usually irregular (lower panels) and probably are not
directly related to the evolution of type~II Cepheids inside the
instability strip. Our observations may indicate that the observed
amplitude and period variations are a manifestation of deterministic
chaos (\eg Buchler \etal 1996, Smolec 2016).

\begin{figure}[h]
\begin{center}
\includegraphics[width=12.7cm]{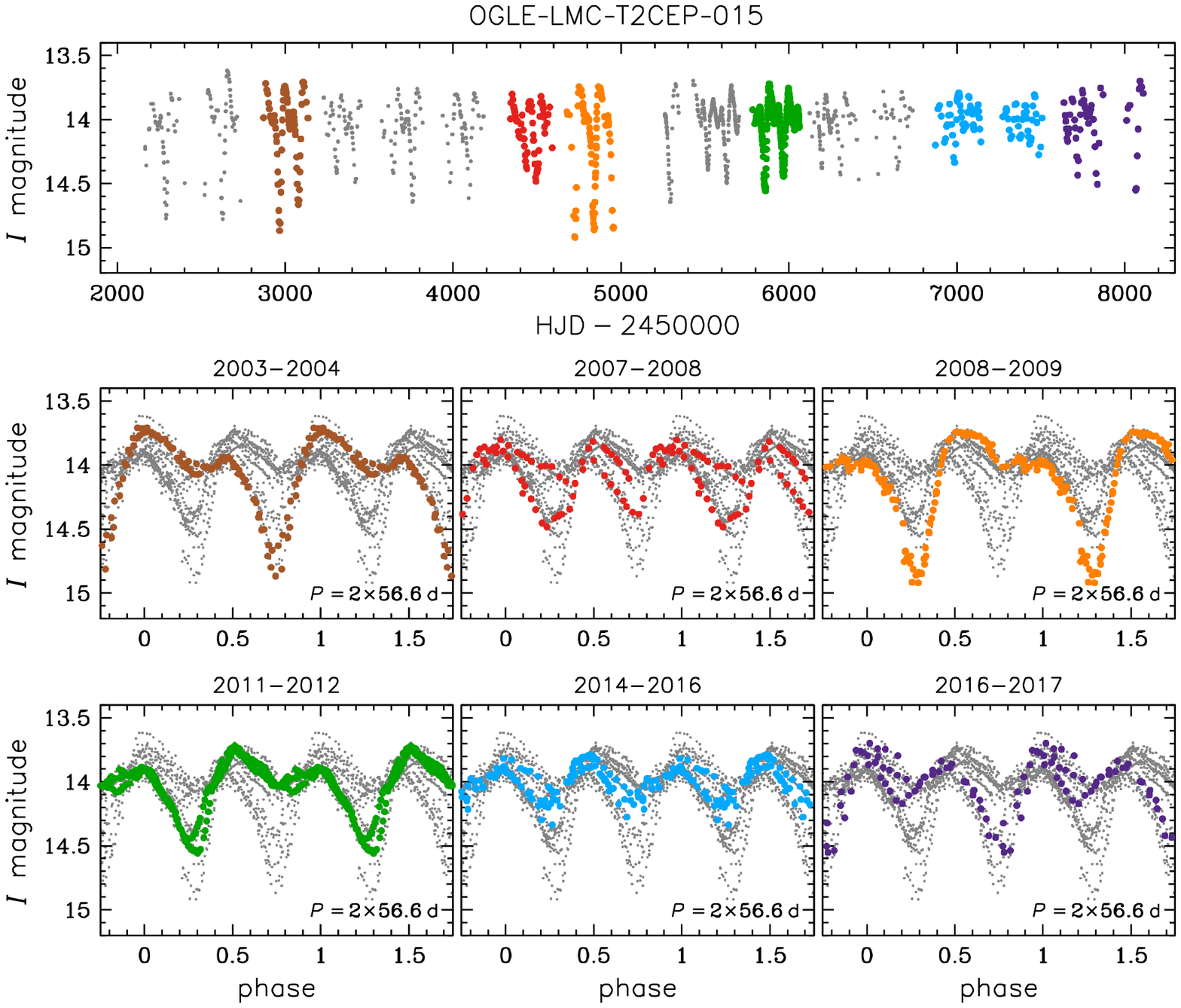}
\end{center}
\FigCap{OGLE-III and OGLE-IV {\it I}-band light curve of
OGLE-LMC-T2CEP-015 -- an RV~Tau star that experienced two interchanges
of deep and shallow minima during 16 years of the OGLE
monitoring. {\it Upper panel} shows the unfolded light curve, {\it
lower panels} present folded light curves. Different colors indicates
observations obtained in selected observing seasons. Note the
interchanges of deep and shallow minima that occurred in years
2004--2008 and 2012--2016.}
\end{figure}

The low-dimensional chaotic behavior may also be responsible for
occasional interchanges of deep and shallow minima observed in some
RV~Tau stars (Plachy \etal 2014). Fig.~10 presents the OGLE light
curve of OGLE-LMC-T2CEP-015, which experienced two interchanges during
16 years when it was monitored by OGLE. The lower panels of Fig.~10
display folded light curves from selected observing seasons, marked
with the same colors in the upper panel. The order of the deep and
shallow minima reversed between 2004 and 2008 and returned to the
previous state in 2016. Both interchanges were separated by episodes
of smaller amplitudes and nearly equal depths of minima, during which
OGLE-LMC-T2CEP-015 might have been classified as a yellow semiregular
variable (SRd star).

\Section{Conclusions}
We presented the most extensive collection of type~II Cepheids in the
Magellanic System. Our sample consists of 338 carefully selected
variables which are almost a complete census of type~II Cepheids in
the LMC and SMC. Together with the lists of variable stars, we provide
their high-quality, long-term light curves in the standard {\it I}-
and {\it V} photometric bands, well suited for exploring properties of
individual objects and their environments. The most obvious directions
for further studies include analyzes of the evolutionary status of
type~II Cepheids based on the spatial distribution, calibrations of
the PL relations, spectroscopic observations of the binary systems
with pulsating components, which may resolve the mystery of peculiar
W~Vir stars.

\Acknow{We would like to thank M.~Kubiak, G.~Pietrzyñ\-ski, and
M.~Pawlak, former members of the OGLE team, for their contribution to the
collection of the OGLE photometric data over the past years. We are
grateful for discussions and constructive comments to R.~Smolec.
We thank Z.~Ko³aczkowski and A.~Schwar\-zen\-berg-Czerny for
providing software used in this study.

This work has been supported by the National Science Centre, Poland,
grant MAESTRO no. 2016/22/A/ST9/00009. The OGLE project has received
funding from the Polish National Science Centre grant MAESTRO no.
2014/14/A/ST9/00121.}

\end{document}